\documentstyle[preprint,aps]{revtex}
\begin{document}
\draft
\title{DIRECTED PERCOLATION DEPINNING MODELS: EVOLUTION EQUATIONS}
\author{L. A. Braunstein\cite{email}, R. C. Buceta and N. Giovambattista}
\address{Departamento de F\'{\i}sica, Facultad de Ciencias
Exactas y Naturales,\\ Universidad Nacional de Mar del Plata,
Argentina}
\author{A. D\'{\i}az-S\'anchez}
\address{Departamento de F\'{\i}sica, Universidad de Murcia, E-30071
Murcia, Espa\~na}
\date{\today}
\maketitle

\begin{abstract}
We present the microscopic equation for the  growing interface
with quenched noise for the model first presented by Buldyrev {\sl
et al.} [Phys. Rev. A {\bf 45}, R8313 (1992)]. The evolution
equation for the height, the mean height, and the roughness are
reached in a simple way. The microscopic equation allows us to
express these equations in two contributions: the contact and the
local one. We compare this two contributions with the ones
obtained for the Tang and Leschhorn model [Phys. Rev A {\bf 45},
R8309 (1992)] by Braunstein {\sl et al.} [Physica A {\bf 266}, 
308 (1999)]. Even when the microscopic mechanisms are quiet
different in both model, the two contribution are qualitatively
similar. An interesting result is that the diffusion contribution,
in the Tang and Leschhorn model, and the contact one, in the
Buldyrev model, leads to an increase of the roughness near the
criticality.
\end{abstract}


\section{INTRODUCTION}
\label{sec:intro}

The description of the noise-driven growth on a self-affine
interface far from equilibrium is a challenging problem. The
interface has been characterized through scaling of the
interfacial width $w$ with time $t$ and lateral size $L$. The
result is the determination of two exponents $\beta$ and $\alpha$
called dynamical and roughness exponents, respectively. It is well
known that interfacial width $w\sim L^\alpha$ for $t\gg t^*$ and
$w\sim t^\beta$ for $t\ll t^*$, where $t^*\simeq L^{\alpha/\beta}$
is the saturation time. These properties occur for many models of
surface growth. The values of the exponents leads to the
classification of these models in different universality classes.
Several models, belonging to the same directed percolation
depinning (DPD) universality class, have been introduced to
explain experiments on fluid imbibition in porous media,
roughening in slow combustion of paper, growth of bacterial
colonies, etc.

It is currently accepted that the quenched disorder plays an
essential role in those experiments. The DPD models take into
account the most important features of the experiments
\cite{Buld,Horv}. The two first models were simultaneously
introduced by Buldyrev {\sl et al.} \cite{Buld} and Tang and
Leschhorn \cite{Tang} to explain the fluid imbibition in paper
sheet. Several authors have been focused their attention on
scaling properties and relationships between the dynamical and
statics exponent for these models. The Tang and Leschhorn (TL)
model has been recently reviewed by Braunstein {\sl et al.}
\cite{Brauns1,Brauns2} from a different point of view than the
traditional one. The principal contribution was the restatement of
the Microscopic Equation (ME) for the TL model (see Appendix).
This equation allows the separation into two contributions: the
substratum contribution by local growth and the diffusion one.
They found that the diffusion contribution to the temporal
derivative of the square roughness may be either negative or
positive and that the behavior of this contribution depends on the
pressure $p$. The negative contribution tends to smooth out the
surface, this case dominate for $p>p_c$ (where $p_c=0.461$ is the
critical pressure). The positive contribution enhances the
roughness. At the critical pressure the substratum contribution to
the temporal derivative of the square roughness is practically
constant, but the diffusion contribution is very strong. This last
contribution, has important duties on the power law behavior.

In this paper we focus the attention in the Buldyrev {\sl et al.}
model of DPD. We show that this model presents several qualitative
features of the TL model. We write a ME, starting from the
microscopic rules, for the evolution of the height as a function of
time. The ME allows us the identification of two contributions
that dominates the dynamics of the system, the ``contact'' and the
``local'' one. In this context we study the mean height speed
(MHS) and the temporal derivative of the square interface width
(DSIW) as a function of time. We show that the contact contribution
smooth out the surface for $p$ well above the criticality but
enhances the roughness near the critical value. To our knowledge
the separation into two contributions for all the quantities
studied in the present paper and the important duties of the
contact contribution to the critical power law behavior has never
been studied before. The paper is organized as follows. In
Section~\ref{sec:bul} we write the microscopic equation for the
evolution of height, the mean height and the roughness for the
Buldyrev {\sl et al.} model. We study the MHS, analyzing the
contact and the local contributions. Also, the two contributions
to the DSIW are analyzed. This separation into two contributions
allows us to explain the mechanism of roughening. In
Section~\ref{sec:tl} we compare the Buldyrev {\sl et al.} model
with the TL model. Finally, we conclude with a discussion in
Section~\ref{sec:concl}.

\section{BULDYREV {\it et al.} MODEL}
\label{sec:bul}

\subsection{Microscopic Equation}

The interface growth takes place in a lattice of edge $L$ with
periodic boundary conditions. A random pinning force $g({\bf r})$
uniformly distributed in $[0,1]$ is assign to every cell of the
lattice. For a given pressure $p$, the cells are divided in two
groups, active (free) cells with $g({\bf r})\le p$ and inactive
(blocked) cells with $g({\bf r})>p$. The interface between wet and
dry cells is specified by a set of integer column heights $h_i$
($i=1,\dots,L$). At $t=0$ we start with flat initial conditions,
{\sl i.e.} $h_i=0$ for all $i$. During the growth, a column is
selected at random with probability $1/L$ and the highest dry
active cell, in the chosen column, that is nearest neighbor to a
wet cell is wetted. Afterwards, we wet all the dry cell below it.
In this model, the time unit is defined as one growth  attempt. In
numerical simulations at each growth attempt, the time $t$ is
increased by $\delta t=1/L$. In this way, after $L$ growth
attempts, the time is increased by one unit. In our simulations we
use $L=\;8192$.

We consider the evolution for the height of the $i$-th site of the
process described above. Let us denote by $h_i(t)$ the height of
the $i$-th generic site at time $t$. From the microscopic rules we
obtain the evolution for the $i$-th height in the next time step
$\delta t = 1/L$
\begin{equation}
h_i(t+\delta t) = h_i(t) + \delta t\;\{
\Theta(-z_i)\,F_i(h_i+1)+[1-\Theta(-z_i)]\,Y_i\}\label{first}\;,
\end{equation}
where $\Theta(x)$ is the unit step function defined as
$\Theta(x)=1$ for $x\ge 0$ and equals to $0$ otherwise,
$F_i(h_i+j)$  equal to 1 if the cell is active and 0 if the cell
is inactive $(1 \le j \le z_i)$, $z_i=\max (h_{i-1},h_{i+1}) -
h_i$, and
\begin{eqnarray*}
Y_i &=& z_i\,F_i(h_i+z_i)\\ &&
+\;(z_i-1)\,F_i(h_i+z_i-1)(1-F_i(h_i+z_i))\\ &&
+\;(z_i-2)\,F_i(h_i+z_i-2)(1-F_i(h_i+z_i-1)(1-F_i(h_i+z_i))+\;\cdots\\
&& +\;F_i(h_i+1)(1-F_i(h_i+2))\dots
(1-F_i(h_i+z_i-1)(1-F_i(h_i+z_i))\;,
\end{eqnarray*}
is the increase of the height in the $i$-th column due to the
contribution of the nearest lateral neighbor. The term between
braces in Eq.~(\ref{first}) takes into account all the possible
ways the site $i$ can growth. The height in the site $i$ is
increased by
\begin{enumerate}
\item\hspace{.5cm}
$1$ \hspace{1.2cm} if $h_i \ge \max(h_{i+1},h_{i-1})$ and
$F_i(h_i+1)=1$,
\item\hspace{.5cm}
$Y_i$ \hspace{1cm} if $h_i < \max(h_{i+1},h_{i-1})$ and
$F_i(h_i+Y_i)=1$.
\end{enumerate}
Otherwise, the height is not increased. We shall call contact
contribution to the term $[1-\Theta(-z_i)]\,Y_i$ (related to case
2) and local contribution to the term $\Theta(-z_i)\,F_i(h_i+1)$
(related to case 1).

Averaging Eq.~(\ref{first}) over the lattice, taking $\delta t\to
0$ the evolution equation for the mean height is
\begin{equation}
\frac{d h}{d t}= \langle \Theta(-z_i)\,F_i\rangle + \langle
(1-\Theta(-z_i))\,Y_i\rangle\;,
\end{equation}
and the evolution equation for the square interface width is
\begin{equation}
\frac{d w^2}{d t}= 2 \langle (h_i-\langle h_i\rangle)
\Theta(-z_i)\,F_i\rangle + 2 \langle (h_i-\langle h_i\rangle)
(1-\Theta(-z_i))\,Y_i\rangle\;.
\end{equation}
The first terms of both equations can be identified as the local
growth contribution, and the second term as the contact growth
contribution. We shall see in Subsection~\ref{subsec:rough} that
the separation into these two analytical terms allows us to show
how the contact mechanism enhances the roughness near the
criticality. In the present paper we focus only on the dynamical
behavior, {\sl i.e.} $t \ll t^* \simeq L $  for the mean height
and roughness (in the DPD models $\alpha/\beta=1$).

\subsection{Mean Height}
\label{subsec:mhs}

The top plot of Figure~\ref{dhdt1} shows the 
MHS as a function of time in three regimes (moving, critical and
pinning phases). The initial condition for the MHS is $p$ in all
regimes. At the criticality we found for the mean height a power
law behavior with the same dynamical exponent that the roughness
one $\beta=0.68\pm 0.02$ for $p_c=\;0.531$. In the moving and
pinning phase we can see that this power law does not hold. Bellow
the criticality, in the pinning phase, the MHS go to zero. Above,
in the moving phase, the MHS goes to certain constant value. The
left plots of Figure \ref{dhdt2} show the contributions to the
MHS: the local one $\langle \Theta(-z_i)\,F_i\rangle$ and the
contact one $\langle (1-\Theta(-z_i))\,Y_i\rangle$. The local
contribution, which is equal to $p$ at $t=0$, is stronger in the
early time regime. This is because in this regime the difference
of heights between nearest neighbors is mostly less than one. This
contribution set into motion the contact growth. In the moving
phase (see left top plot of Figure \ref{dhdt2}) both contributions
go to a certain constant. In the intermediate regime the local
contribution decreases while the contact one increases. At the
criticality and in the pinning phase (see left middle and
left bottom plots of Figure \ref{dhdt2}, respectively) the local
contribution decreases continuously from $p$ to zero. The contact
contribution increases from zero to a maximum value and then
decreases reaching asymptotically the MHS. In all phases both
contributions are equal at $t\simeq 1$. This means that after
$L$-growth attempts the interplay between both mechanisms are
equal independently of $p$. After this, the dynamical behavior is
strongly dominated by the contact mechanism.

\subsection{Roughness}
\label{subsec:rough}

The top plot of Figure \ref{dw2dt} shows the temporal 
(DSIW) as a function of time for
various values of $p$. The initial condition is $p$ in all
regimes. As we expected \cite{Brauns3}, the power law holds only
at the criticality. The DSIW goes asymptotically to zero at the
pinning and moving phase. In the left plots of Figure
\ref{dw2dt2}, we show the two contributions to the DSIW for
different values of $p$. The local contribution $2 \langle
(h_i-\langle h_i\rangle) \Theta(-z_i)\,F_i\rangle$ to the DSIW is
always positive. As $p$ decreases, this contribution becomes less
important, but always rough the interface. On the other hand, for
$p>p_c$, the contact contribution $2 \langle (h_i-\langle
h_i\rangle) (1-\Theta(-z_i))\,Y_i\rangle$ can take negative
values, smoothing out the surface. Otherwise, for $p\le p_c$, the
contact contribution is always positive roughening the interface.
One could expect that the contact contribution always smooth out
the surface because it tends to widen the roughen picks. However,
near the criticality, the contact growth happens mainly in lateral
neighbors cells to few height terraces above the mean height.
Then, this new wetted column smooth out locally, but it moves away
from the mean height increasing the roughness.

\section{COMPARISONS WITH THE TANG AND LESCHHORN MODEL}
\label{sec:tl}

We rescue the similarities between the Buldyrev {\sl et al.} and
the Tang and Leschhorn models despite the strong microscopic
differences between their rules. In a previous paper Braunstein
{\sl et al.} \cite{Brauns1,Brauns2} wrote the ME for the TL model.
They identified two separate contributions: the substratum and the
diffusion one in the MHS and the DSIW (see Appendix). In the
present paper, we also obtain two contributions: the local and the
contact one. Figure \ref{dhdt2} shows the contributions to the MHS
for both models. Notice that each pair of plots have the same
qualitatively behavior. The shape of the diffusion and substratum
contribution in the TL model are qualitatively similar with the
contact and local contribution in the Buldyrev {\sl et al.} model,
respectively, even when the microscopic processes are quite
different for each model. The different contributions to the DSIW
for both models are shown in Figure \ref{dw2dt2}. Notice that the
diffusion and the contact contributions, in each model, can take
negative values for $p>p_c$ smoothing out the interface. Near the
criticality, in both models, the roughness is mainly due to the
diffusion and the contact contributions. These last contributions
play a very important role at the criticality in each model. The
similarities between the models could explain why these two
different microscopic models belongs to the same universality
class. Figure \ref{dw2dt} shows the DSIW as a function of time for
both models. As we expected \cite{Brauns3}, the power law holds
only at the criticality despite other authors
\cite{Buld,Tang,Hernan,Yang,Amaral}. From these plots it is easy
to see that this last statement holds for both models.

\section{CONCLUSIONS}
\label{sec:concl}

We wrote the ME for the evolution of the height in the Buldyrev
{\sl et al.} model and we compared the results obtained with those
from the TL model. Using the ME we studied the evolution of the
mean height and the roughness. The ME allows us to separate the
local and the contact contributions. We found that the contact
contribution near the criticality is the main responsibility of the
roughness. We found qualitatively that the shape of the contact
contribution is analogous to that of the diffusion one in the TL
model, and that the shape of the local growth is similar to that
of the substratum contribution in the TL model. We found that the
power law behavior holds only at the criticality for the Buldyrev
{\sl et al.} model. This last feature, common to the TL model,
suggests to us that it may be common to all other DPD growth models
\cite{Amaral}.

\acknowledgements

L. A. Braunstein acknowledge the financial support
from FOMEC 290, Argentina.

\appendix
\section*{MICROSCOPIC EQUATION FOR TANG AND LESCHHORN MODEL}

We present here the microscopic equation for the TL model
\cite{Brauns1,Brauns2}.
The time evolution equation for the interface height, in a time step
$\delta t$, is
\begin{equation}
h_i(t+\delta t) = h_i(t)+\delta t\; [W_{i+1}+W_{i-1}+F_i(h_i')\,W_i]
\;,\label{me}
\end{equation}
with
\begin{eqnarray*}
W_{i}&=&1-\Theta(h_i-\min(h_{i-1},h_{i+1}) - 2)\nonumber\;,\\
W_{i\pm 1}&=&\Theta(h_{i\pm 1}-\min(h_i,h_{i\pm 2})-2)
\{ [1-\Theta(h_i-h_{i\pm 2})]+\case{1}{2}
\delta_{h_i,h_{i\pm 2}}\} \;.\nonumber\\
\end{eqnarray*}
where $h_i'=h_i+1$. Here $\Theta(x)$ is the unit step function defined
as $\Theta(x)=1$ for $x\ge 0$ and equals to $0$ otherwise. $F_i(h_i')$
equals to $1$ if the cell at the height $h_i'$ is free or active
({\sl i.e.} the growth may occur at the next step) or $0$ if the cell is
blocked or inactive. $F_i$ is the interface activity function.

Averaging over the lattice, taking $\delta t\to 0$
the evolution equation for the mean height is
\begin{equation}
\frac{dh}{dt}=
\langle W_i\,F_i\rangle + \langle 1-W_i \rangle \label{dhdt}\;,
\end{equation}
and the evolution equation for the square interface width is
\begin{equation}
\frac{d w^2}{d t}= 2 \langle (h_i-\langle h_i\rangle) W_i\,F_i\rangle
+ 2 \langle [\min (h_{i-1},h_{i+1})-\langle h_i\rangle] (1-W_i)\rangle\;.
\end{equation}
The first terms of both equations have been identified as the substratum growth
contributions, and the second terms as the diffusion growth contribution.

\newpage

\begin{figure}
\caption{Plots of $p^{-1}\,dh/dt$ vs $t$. The top (bottom) plot
shows the results for the Buldyrev {\sl et al.} (Tang and
Leschhorn) model. For the top plot the parameter $p$ is 0.56
($\bigtriangleup$), 0.531 ($\bigcirc$) and 0.51
($\bigtriangledown$). For the bottom plot the parameter $p$ is
0.49 ($\bigtriangleup$), 0.461 ($\bigcirc$) and 0.4
($\bigtriangledown$).} \label{dhdt1}
\end{figure}

\begin{figure}
\caption{ln-ln plots of different contributions to the MHS as a
function of time for the Buldyrev {\sl et al.} model (left plots)
and the Tang and Leschhorn model (right plots), for different
values of $p$. The circles represent the contact contribution
(left plots) and the diffusion contribution (right plots). The
squares represent the local contribution (left plots) and the
substratum contribution (right plots). For both models, the top,
bottom and middle plots show the moving, the pinning and the
critical phases, respectively.} \label{dhdt2}
\end{figure}

\begin{figure}
\caption{DSIW as a function of time. The top (bottom) plot shows the
results for the Buldyrev {\sl et al.} (Tang and Leschhorn) model.
For the top plot the parameter $p$ is 0.7 ($\bigcirc$), 0.56
($\bigtriangledown$), 0.531 ($\bullet$) and 0.51 ($\bigtriangleup$).
For the bottom plot the parameter $p$ is 0.7 ($\bigcirc$), 0.49
($\bigtriangledown$), 0.461 ($\bullet$) and 0.4 ($\bigtriangleup$).
In both models the symbol $\bullet$ shows the critical behavior.}
\label{dw2dt}
\end{figure}

\begin{figure}
\caption{Semi-ln plots of different contributions to the DSIW as
function of time for the Buldyrev {\sl et al.} model (left plots)
and the Tang and Leschhorn model (right plots), for different
values of $p$. The circles represent the contact contribution
(left plots) and the diffusion contribution (right plots). The
squares represent the local contribution (left plots) and the
substratum contribution (right plots). For both models, the top,
bottom and middle plots show the moving, the pinning and the
critical regime, respectively.} \label{dw2dt2}
\end{figure}

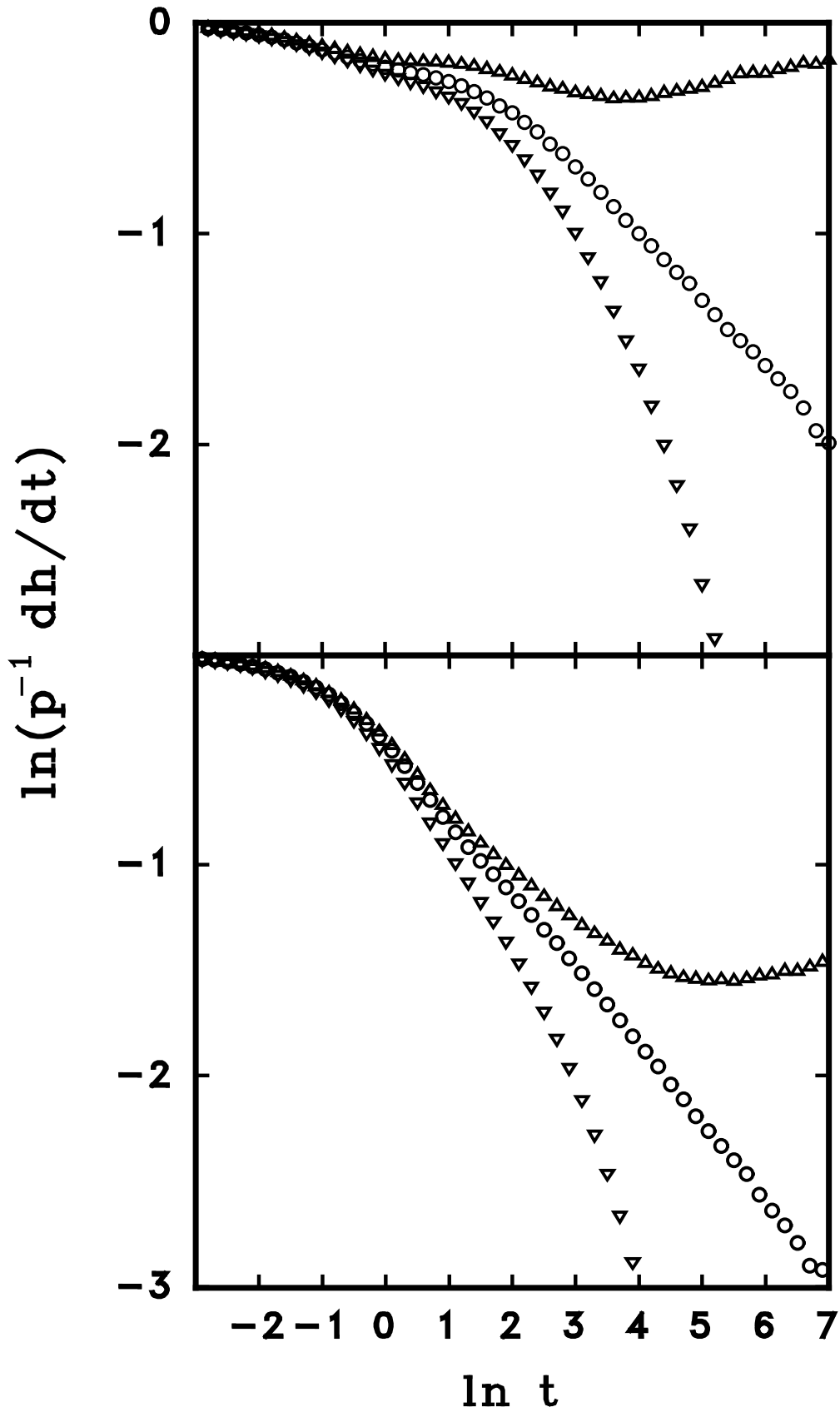
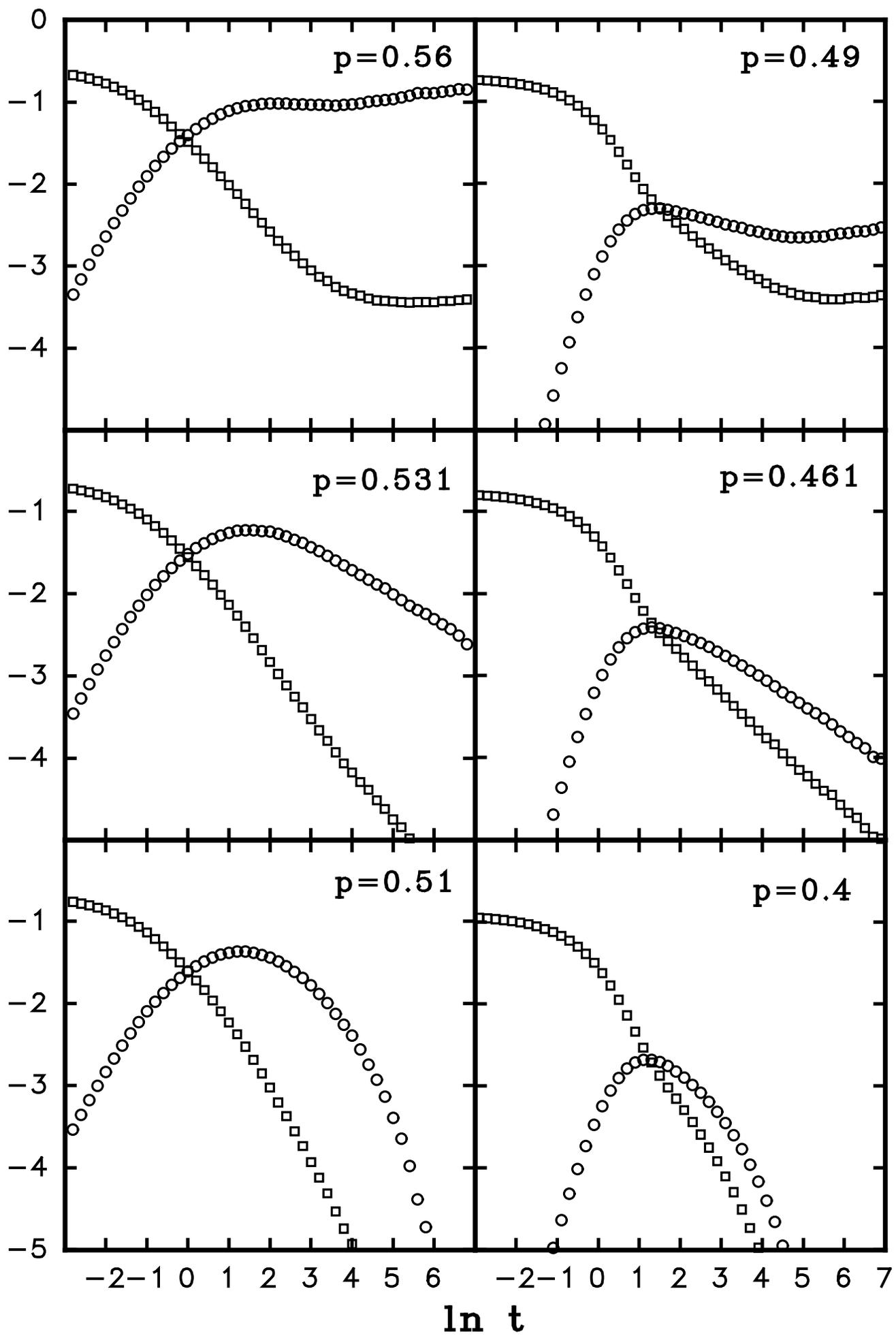
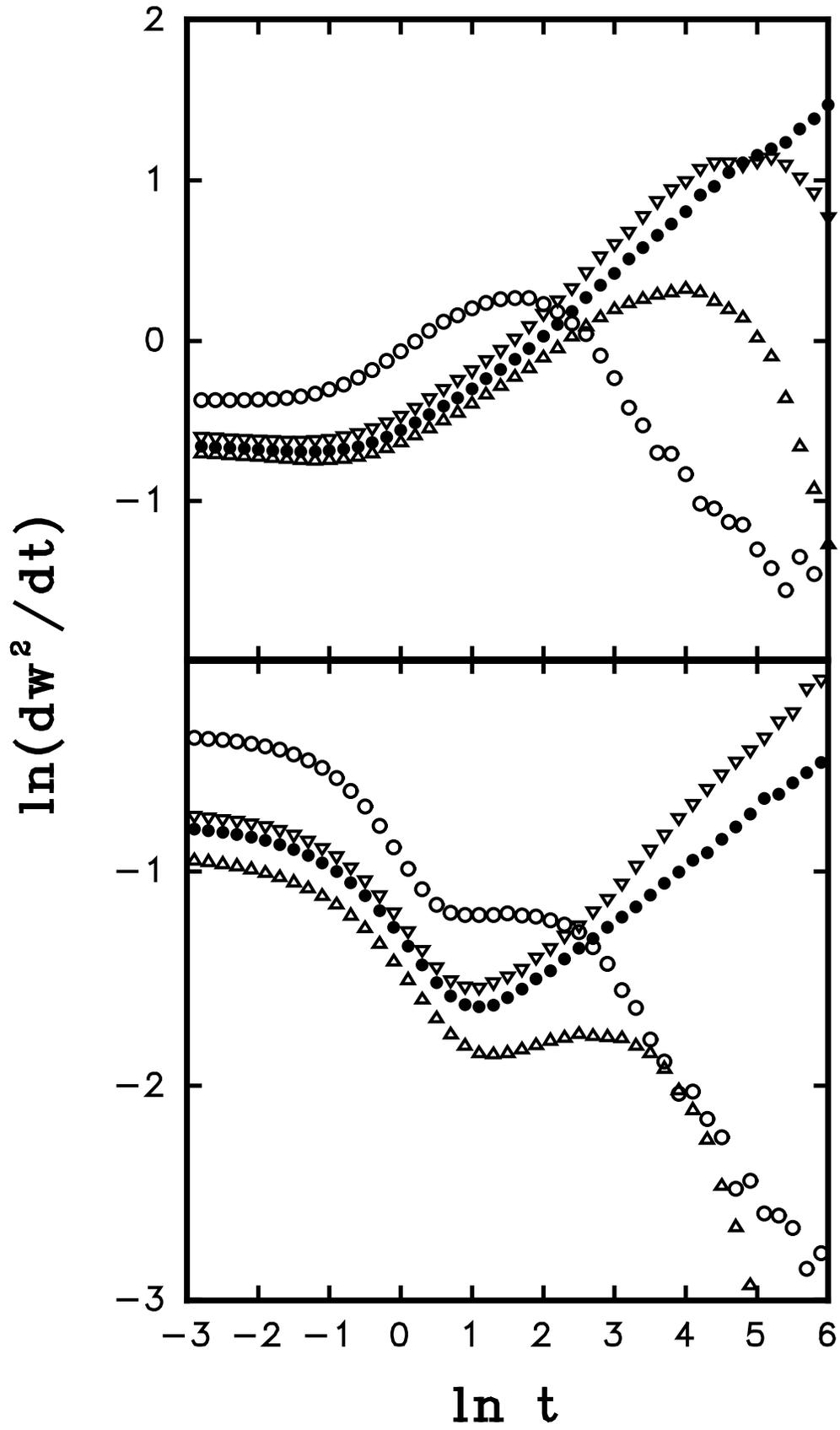
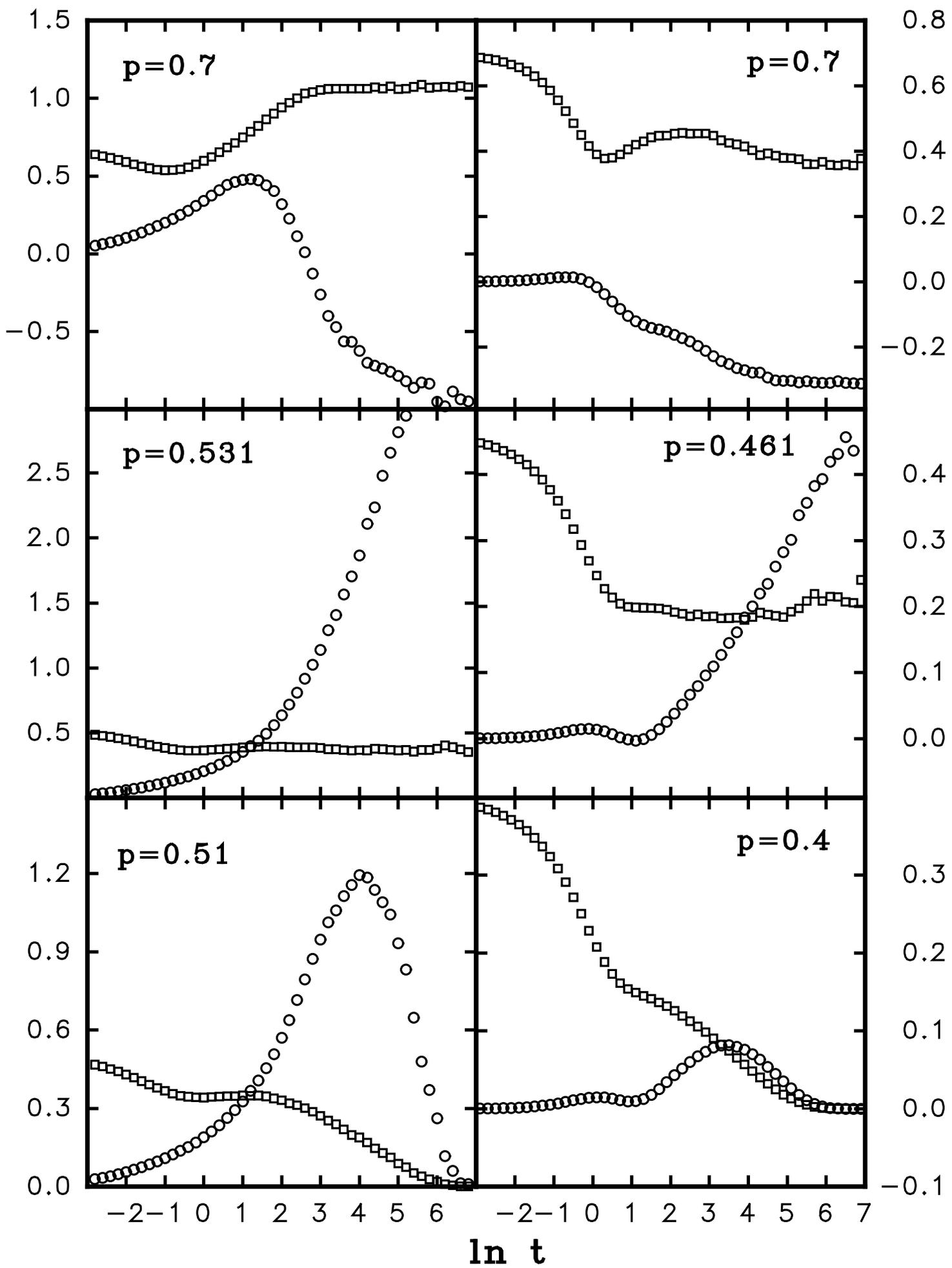
\end{document}